\documentclass[amsmath,amssymb,aps,prb,10pt,twocolumn, longbibliography]{revtex4-2}
\usepackage{silence}
\newcommand{\tsat}{{\textsc{Tsat}}}
\newcommand{\walksat}{\textsc{Walksat}}
\newcommand{\docsat}{\textsc{Docsat}}
\newcommand{\cadical}{\textsc{CaDiCaL}}
\newcommand{\yalsat}{\textsc{Yalsat}}
\newcommand{\surveyprop}{\textsc{SP}}

\newcommand{\Title}{Targeting Clause Type Distributions: a Picklock for Random Satisfiability Problems}

\usepackage[acronym]{glossaries}
\newacronym{sls}{SLS}{stochastic local search}
\newacronym{tlc}{TLC}{true literal count}
\newacronym{ct}{CT}{clause-type}
\newacronym{ctd}{CTD}{clause-type distribution}
\newacronym{ctds}{CTDS}{clause-type distribution space}
\newacronym{cnf}{CNF}{conjunctive normal form}

\makeatletter
\newcommand*{\glsplainhyperlink}[2]{%
    \begingroup%
      \hypersetup{hidelinks}%
      \hyperlink{#1}{#2}%
    \endgroup%
}
\let\@glslink\glsplainhyperlink
\makeatother

\let\label\undefined
\usepackage{hyperref, nameref}

\usepackage{algorithm}
\usepackage{algpseudocode}
\usepackage{xcolor}
\usepackage{graphicx}
\usepackage{textcomp, gensymb}
\usepackage{bm, bbm} % bold math symbols, e.g. \bm{\sigma} etc.
\usepackage{csquotes}
\usepackage[all]{nowidow}

\DeclareMathOperator{\sgn}{sgn}

\newcommand{\algoref}[1]{\hyperref[#1]{Algorithm~\ref*{#1}}}
\newcommand{\appref}[1]{\hyperref[#1]{Appendix~\ref*{#1}}}

\newcommand{\jcb}[1]{\bgroup\color{orange}JCB: #1\egroup}
\newcommand{\jsc}[1]{\bgroup\color{purple!75!blue}JS: #1\egroup}
\newcommand{\TODO}[1]{\bgroup\color{red!60!black}TODO: #1\egroup}
\newcommand{\modified}[1]{\bgroup\color{red!80!black}#1\egroup}

\begin{document}

\title{\Title}
\author{Joachim Schwardt$^{1,2}$}
\email{jschwardt@pks.mpg.de}
\author{Jan Carl Budich$^{1,2}$}

\affiliation{$^1$Max Planck Institute for the Physics of Complex Systems, N\"{o}thnitzer Str.~38, 01187 Dresden, Germany}
\affiliation{$^2$Institute of Theoretical Physics, Technische Universit\"{a}t Dresden and W\"{u}rzburg-Dresden Cluster of Excellence ctd.qmat, 01062 Dresden, Germany}

\date{\today}

\begin{abstract}
Optimization problems such as the NP-complete 3-SAT provide an important benchmark for the difficult task of finding ground-states in strongly correlated many-body systems with rugged energy landscapes.
The study of random 3-SAT problems as Ising spin Hamiltonians in statistical physics has yielded major insights including the existence of a satisfiability phase transition, and the prediction of a critical parameter line of particularly hard instances.
Yet, progress on solving those instances has been scarce for several decades.
Here, introducing the \textsc{Target-SAT} (\tsat{}) algorithm, we roughly triple the tractable problem sizes in the hardest regime, with an even greater improvement in a vast range of neighboring regions.
By leveraging statistical information hidden in the combinatorial constraints of the problem, \tsat{} is actively guided in its stochastic local search toward a target within the relevant parameter space.
Our analysis also explains why established local search algorithms are limited to relatively small system sizes due to a vast low-energy trap.
Furthermore, we characterize the aforementioned critical line in terms of a dominant additional complexity barrier, whose exponential scaling is quickly overcome by \tsat{} only in the surrounding parameter space.
With \tsat{}, the lead in solving the hardest known random satisfiability problems returns to the realm of stochastic local search algorithms.
\glsresetall
\end{abstract}

\maketitle

\section{Introduction}\label{sec:intro}
Finding the ground-state of a strongly correlated many-body system is among the foundational challenges in physics, which may be seen as a hard optimization problem arising from the emergence of local minima and barriers in rugged energy landscapes.
Physically, such obstacles may for example be rooted in frustrated magnetic interactions, that amount to competing combinatorial constraints from a mathematical perspective.
Owing to this close correspondence, mappings between Ising spin models in physics and abstract satisifiablity problems \cite{np_complete.comp_complexity.ising_spinglass.Barahona_1982,np_complete.ising_formulations.lucas_2014} have stimulated an active frontier of interdisciplinary research for decades, encompassing statistical physics \cite{spinglass.sk_model.infinite_order_pars.Parisi_1979,spinglass.sk_model.solution_sequences.Parisi_1980,eo.algorithm.Boettcher_2001,ea.spinglass.perf_match.Bieche_1980,ea.spinglass.perf_match.Kasteleyn_cities.Thomas_2007,optimization.algorithms.physics.Hartmann_2006}, computational complexity theory \cite{sat.NP_complete.Cook_1971,np_complete.reducibility.combinatoric_problems.Karp_1972,np_complete.guide.computers_intractibility.Garey_1990,sat.pt.random.two_moments.Achlioptas_2006,sat.np.complexity.eth.Impagliazzo_2001}, and constrained optimization \cite{sat.qubo.comb.optimization.landscapes.deep_minima.Dobrynin_2024,sat.practical.review.complete.Kullmann_2008,sat.circuits.Nam_1999,sat.circuits.FPGA_islands.Mukherjee_2015,ai.MMP.MarquesSilva_2013,ai.prob_reasoning.maxsat.Park_2002,ai.constraint_processing.Dechter_2003,ai.maxsat.maxsolver.Xing_2005,sat.protein.Ollikainen_2009,sat.protein.optimization.Allouche_2014,sat.model_checking.verification.Gupta_2006,sat.model_checking.survey.Prasad_2005,sat.timetables.Matos_2021,sat.routing.biochips.Yuh_2011,sat.routing.system_on_a_chip.Zhukov_2020}.
In this context, the paradigmatic NP-complete 3-SAT problem has become a ubiquitous benchmark problem around which crucial progress has evolved in various directions.

\begin{figure}[htp!]
    \centering
    \includegraphics{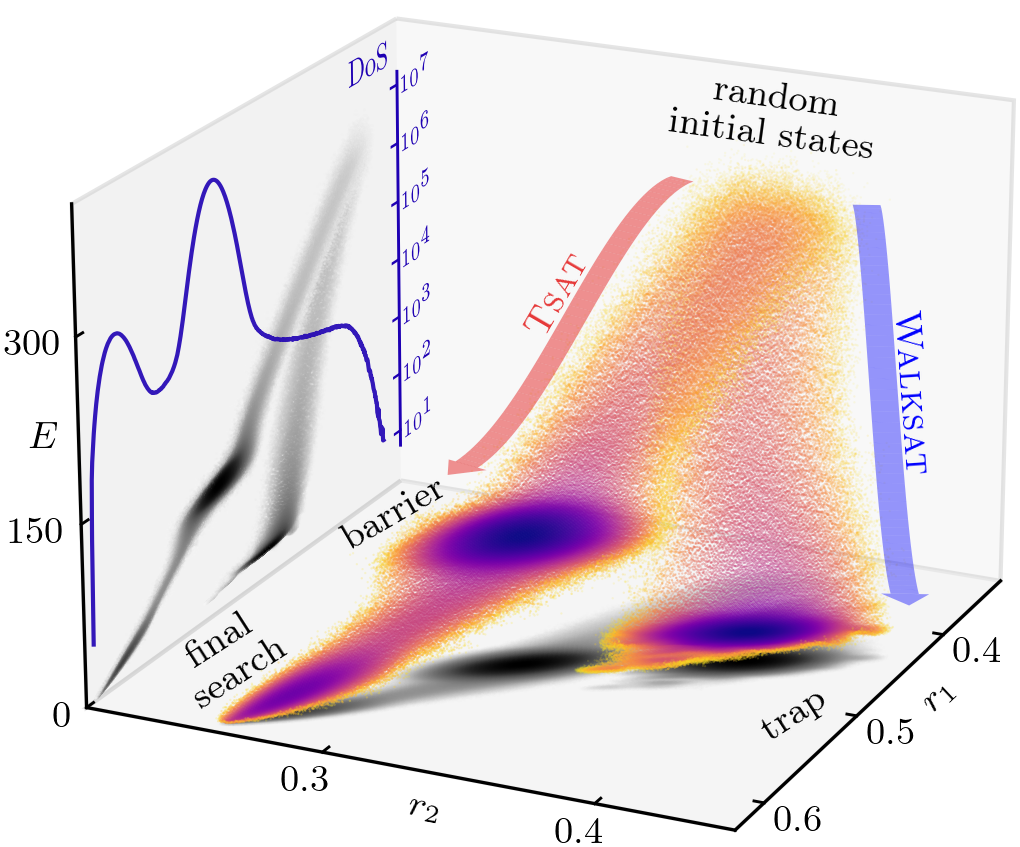}
    \caption{Illustration of \tsat{} and \walksat{}  trajectories through \acrfull{ctds}, starting from random initial states.
    The color corresponds to the $\log_{10}$ of the density of states (DoS), and for the \tsat{} trajectories we also show the accumulated DoS on the left (cf.~\autoref{fig:timescales_ctds}).
    The grayscale projections are along the $E$ and $r_2$ axes and serve as a visual guide.
    \walksat{} does not reach the solution at $E=0$ because it gets stuck in a vast local minimum of low-energy states (trap).
    The \tsat{} algorithm introduced in this article takes a completely different path and encounters a barrier before entering a final search for the solution located near the bottom left.
    }
    \label{fig:caricature_3d}
\end{figure}

For the generic case of random 3-SAT instances, there exists a sharp phase transition to non-zero ground-state energies (i.e. unsatisfiable problems) at a critical density $\alpha_\text{c}$ \cite{sat.hard_regime.Cheeseman_1991,sat.hard_regime.distributions.Mitchell_1992,sat.pt.random.two_moments.Achlioptas_2006,sat.pt.cavity.mc_data.alpha_crit.Lundow_2019} of combinatorial constraints.
An important result from statistical physics is the explanation of this transition via the so-called cavity method at \enquote{one-step replica symmetry breaking level} \cite{sat.pt.cavity.Mertens_2006,sat.pt.cavity.random.Mezard_2002}, which, although some mathematical concerns about its rigor remain \cite{sat.pt.cavity.survey_prop.Coja_2017,sat.pt.cavity.survey_prop.info_theo_threshold.Coja_2017,sat.survey_prop.random.analysis.Hetterich_2016}, matches the numerical observations quite well.
This approach has also shed light on the structure of the low-energy states, and thereby lead to the powerful \textsc{Survey Propagation} (\surveyprop{}) algorithm \cite{sls.sp.Braunstein_2005,sls.sp.bsp.ksat.Marino_2016} capable of finding ground-states for very large systems close to the phase transition.
However, further insights in the physical context of magnetism have revealed a parameter regime for random 3-SAT problems that are beyond the capabilities of \surveyprop{}, and that always have zero-energy ground-states due to an inbuilt symmetry \cite{sat.hard_random.generator.statmech.Weigt_2002}.
In particular, on a \emph{critical line} \footnote{To clarify, the word critical here refers to the critically hard region within the parameter regime of the hidden solution at some fixed clause density $\alpha$, and not to the critical value $\alpha_\text{c}$ that marks the phase transition toward unsatisfiability in random 3-SAT problems.} in the statistical distribution of problem parameters, ground-states have an extensive number of frozen spins (referred to as a \emph{backbone}), whose magnetization is fixed across all ground-states thus allowing them to hide well in the exponentially large configuration space \cite{sat.hard_random.generator.statmech.Weigt_2002}.
For such problems, all known optimization methods, including \gls{sls} \cite{sls.walksat.Selman_1994,SLS.walksat.review.Hoos_2000,SLS.review.overview.Hoos_2015,SLS.WalkSAT.random_3sat.Fu_2020}, and so-called complete solvers \cite{sat.CDCL.solvers.Silva_2009,sat.preprocessing.Een_2005,solvers.cadical2.Biere_2024}, despite their remarkable advances on structured (practical) problems \cite{sat.handbook.Biere_2009,sat.review.advances.Alouneh_2019,sat.revolution.Fichte_2023}, are limited to modest problem sizes ($N \sim 500$), and have seen little progress over several decades.

Here, we present a \gls{sls} solver coined \textsc{Target-SAT} (\tsat{}) that roughly triples the tractable problem sizes to $N \sim 1500$ in the aforementioned hardest regime of 3-SAT, and that yields even greater improvement in an extended adjacent parameter region (cf.~\autoref{fig:psuccess_demo}).
Our findings are based on analyzing and harnessing the statistical properties of combinatorial constraints in random 3-SAT instances, organized in a phase space coined \gls{ctds} (cf.~\autoref{fig:caricature_3d}).
Specifically, drawing inspiration from our recent more specialized heuristic \docsat{} \cite{sls.docsat.Schwardt_2025}, we generally discuss how the position in \gls{ctds} affects the difficulty of a problem, and how targeting this position provides efficient guidance to \gls{sls} beyond energy $E$ as the primary figure of merit.
This approach not only allows us to empirically confirm the predicted \cite{sat.hard_random.generator.statmech.Weigt_2002} critical line as the location of the hardest problems (cf.~\autoref{fig:fullspace_n1000_wsat_tsat}), but also to reveal a deeper structure of their complexity in terms of two separate barriers (cf.~\autoref{fig:caricature_3d}).
Quite remarkably, with our \tsat{} algorithm the first complexity barrier representing the main bottleneck on the critical line (cf.~\autoref{fig:timescales_ctds}) quite quickly becomes sub-exponential when moving away in \gls{ctds} (cf.~\autoref{fig:timescales_ctds_065}).
By contrast, the second complexity barrier dubbed \emph{final search} generically remains exponential so as to reflect the NP-completeness of 3-SAT, and behaves smoothly around the critical line (cf.~\autoref{fig:timescales_ctds_avg_tlc_line}).
Moreover, our study clarifies why well established \gls{sls} algorithms such as \walksat{} \cite{sls.walksat.Selman_1994} inevitably run into a trap of local minima (cf.~\autoref{fig:caricature_3d}) as they do not deliberately target specific regions in \gls{ctds}.
Outperforming all other methods in a wide parameter range including the hardest solvable 3-SAT problems, \tsat{} drastically advances and reclaims for \gls{sls} algorithms the forefront of solving random satisfiability problems.

\section{Clause-Type Distributions in 3-SAT}\label{sec:ctds}
In this section, we establish the nomenclature of the 3-SAT problem as well as the concept of clause-types, and describe the mapping to an Ising Hamiltonian.
We briefly discuss \gls{sls} algorithms exemplified by the \walksat{} heuristic, and then show how their performance is intimately linked to the aforementioned statistics of the problem instances.

\subsection{From combinatorics to Ising Hamiltonians}
In combinatorial optimization, a 3-SAT problem consists of $N$ Boolean \emph{variables} $\bm{x}=(x_1,\dots,x_N)$.
The $x_n$ are constrained by $M = \alpha N$ clauses, with the \emph{clause density} $\alpha$.
Each clause $C_m=l_{m_1}\lor l_{m_2} \lor l_{m_3}$ connects three \emph{literals} via a logical or, where each literal $l$ can represent either a variable $x$ or its negation $\bar{x}$.
A full 3-SAT problem is simply the conjunction of all clauses $C_1 \land \ldots \land C_M$, referred to as \gls{cnf}, requiring that all clauses must be satisfied simultaneously.

In statistical physics, this problem may be equivalently represented as an Ising spin Hamiltonian 
\begin{align}
    H &= \sum_{m=1}^M \tilde{C}_m,
\end{align}
where the terms $\tilde{C}_m=P(l_{m_1})P(l_{m_2})P(l_{m_3})$ are defined with the projectors $P(l_n) = \frac{1 - \sgn(l_n) \sigma_{n}}{2}$.
Here, a negative sign of $l_n$ corresponds to a negated variable $\bar{x}_n$.
Physically we may think of $\sigma_n \in \{\pm 1\}$ as spins with $\sigma_n=2x_n-1$. % $\sigma_n=+1$ for $x_n=1$ and $\sigma_n=-1$ for $x_n=0$.
Each clause has an \emph{energy} $E$ of zero if it is satisfied and an energy of one otherwise.
For a state $\bm{x}$, the total energy is then simply defined as
\begin{align}
    E(\bm{x}) &= \text{number of clauses violated by }\bm{x}, \label{eqn:energy}
\end{align}
such that a ground-state $\bm{s}$ (or \emph{solution}) must satisfy all clauses and therefore have $E(\bm{s})=0$.
In the spin representation, the problem is equivalent to $H\bm{\sigma}=E\bm{\sigma}$, where solutions are zero-energy eigenvectors of $H$.

\subsection{Clause type distribution space}
On a more detailed note, each clause represents a constraint in the space of $2^N$ possible states, and e.g. $C_1=\bar{x}_2\lor x_4 \lor \bar{x}_5$ is satisfied by 7 out of the $2^3=8$ possible assignments to the involved triplet of variables.
For one of these, say $x_2=0$, $x_4=0$ and $x_5=1$, the individual literals of the clause $C_1$ become either \emph{true} ($\bar{x}_2=1$) or \emph{false} ($x_4=0$, $\bar{x}_5=0$).
We say that a clause is type-$k$ if it has exactly $k$ true literals, and we define 
\begin{align}
    m_k(\bm{x}) = \text{number of type-$k$ clauses given }\bm{x}, \label{eqn:m_k_of_x}
\end{align}
noting that $m_0 \equiv E$ is the number of unsatisfied clauses.
We are usually more interested in the relative number of clause-types and therefore introduce the ratios
\begin{align}
    r_k(\bm{x}) &= \frac{m_k(\bm{x})}{M} \in [0,1]. \label{eqn:r_k_of_x}
\end{align}
We refer to $\bm{r}=(r_0,\dots,r_3)$ as the \gls{ctd} of a state $\bm{x}$, which admits the normalization $\sum_{k=0}^3 r_k(\bm{x}) = 1$, and we refer to the set of all possible $\bm{r}$ as the \acrfull{ctds}.
Note that because a solution $\bm{s}$ has $r_0=0$, i.e. no unsatisfied (type-0) clauses, the normalization implies that the two numbers $r_1(\bm{s})$ and $r_2(\bm{s})$ are sufficient to parameterize the solution-subspace of the full \gls{ctds} of 3-SAT.

\subsection{Stochastic local search}
\Acrfull{sls} solvers generally try to solve a satisfiability problem via a focused search starting from a random initial state $\bm{x}$, iteratively flipping the value of variables chosen by a heuristic \cite{sls.walksat.Selman_1994,SLS.walksat.review.Hoos_2000,SLS.review.overview.Hoos_2015,SLS.WalkSAT.random_3sat.Fu_2020}.
Focused means that every iteration of the algorithm starts by (randomly) selecting one of the violated constraints (type-0 clauses), which has the benefit of satisfying the chosen clause regardless of which of its variables is selected.
This process is iterated for a certain number of flips, after which the search is usually restarted from a new random state.
The key ingredient of such \gls{sls} algorithms then lies in their heuristic selection process of variables and clauses.

With the \emph{walk probability} $p_\text{walk}$, the well established \gls{sls} heuristic \walksat{} picks a random variable from a randomly chosen type-0 clause, while with $1-p_\text{walk}$ it instead performs a greedy step by picking the variable with minimal breakcount $b$ in that clause \cite{sls.walksat.Selman_1994}.
The breakcount $b_n$ for a variable $x_n$ is the number of clauses that would be broken by flipping $x_n$ (i.e. the number of transitions from type-1 to type-0 clauses).
For completeness, we note that \walksat{} always picks a variable with $b_n=0$ if it exists in the selected clause (i.e. \enquote{free} moves are preferred), and that any ties are broken randomly.

\subsection{Satisfiable random problems with fixed CTD}
We now clarify where the hardest 3-SAT problems are expected to be found for a given problem size $N$.
In \emph{random uniform} instances, clauses are generated by choosing three variables out of the available $N$ as literals, and to negate each of them independently with probability $\frac{1}{2}$.
Finding solutions to such problems is comparably well under control.
More specifically, they are known to undergo a phase transition at a critical clause density of $\alpha_\text{c}\approx 4.262$ \cite{sat.pt.cavity.mc_data.alpha_crit.Lundow_2019}, beyond which instances are unsatisfiable with high probability \cite{sat.hard_regime.Cheeseman_1991,sat.hard_regime.distributions.Mitchell_1992,sat.pt.random.two_moments.Achlioptas_2006}.
Below $\alpha_\text{c}$, the solutions are well understood theoretically via the cavity method \cite{sat.pt.cavity.Mertens_2006,sat.pt.cavity.random.Mezard_2002} and also exploitable computationally via the \textsc{Survey Propagation} (\surveyprop{}) algorithm derived from it \cite{sls.sp.Braunstein_2005,sls.sp.bsp.ksat.Marino_2016}.
As $\alpha$ approaches the phase transition, the initially connected solution space shatters into exponentially many small clusters, which eventually vanish at $\alpha_\text{c}$.
We refer to such solutions as \emph{accidental}, because they originate from a lack of constraints and always exist for sufficiently small $\alpha < \alpha_\text{c}$ unless there is additional structure in the problem that prevents their formation (e.g. finite interaction ranges).

Since \surveyprop{} can find these states even beyond $N=10^5$, $\alpha < \alpha_\text{c}$ is clearly not \enquote{as hard as it gets} for 3-SAT and we therefore require $\alpha\gtrsim \alpha_\text{c}$.
To ensure  existence of zero-energy ground-states for larger $\alpha$, modified (non-uniform) protocols for generating hard solvable instances have been introduced  \cite{sat.hard_random.generator.statmech.Weigt_2002}.
Such problems have a hidden solution at a predetermined \gls{ctd} (i.e. fixed $r_1$ and $r_2$; cf. \appref{app:cnf}), and a \emph{critical} \cite{Note1} parameter-line has been predicted for which finding the solution is hardest \cite{sat.hard_random.generator.statmech.Weigt_2002}.
Here, we study more broadly how $r_1$ and $r_2$ affect the problem difficulty.
To suppress the coexistence probability of usually easier to find accidental solutions, we consistently choose $\alpha > \alpha_\text{c}$ by a finite margin in the following.
Specifically, to mitigate finite-size effects at moderate $N$, we set $\alpha=5$ so as to avoid accidental solutions entirely for all studied system sizes.
We note that the higher number of constraints may make the problems \enquote{slightly less hard} \cite{sat.complexity.around_alpha_c.Coarfa_2000} for complete solvers such as \cadical{} \cite{solvers.cadical2.Biere_2024}, but these differences are minor close enough to $\alpha_\text{c}$.
In this light, we emphasize that $\alpha = 5$ may make the performance comparison in \appref{app:comparison_others} somewhat unfavorable for our \gls{sls} algorithm, but improves the clarity of the numerical results for finite $N$, and is thus a conservative choice for our subsequent benchmarks.

\begin{figure}[!htp]
    \centering
    \includegraphics{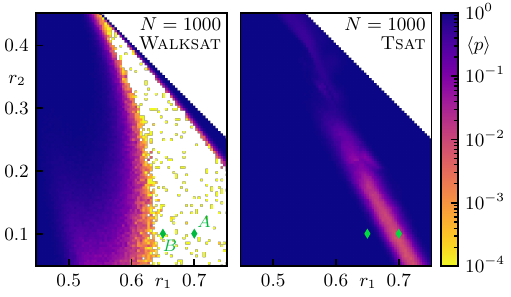}
    \caption{Average success probability $\langle p \rangle$ of \walksat{} (left panel) and \tsat{} (right panel) for problems generated via \algoref{alg:cnf}.
    At each point in \gls{ctds} we generate $10^2$ instances with $N=1000$ and $\alpha=5$, using $10^2$ trials per instance and $300N$ flips per trial (cf.~\appref{app:parameters} for solver specific parameters).
    Even on the clearly distinguished critical line $r_2=\frac{3}{2}-2r_1$ predicted to harbor the most difficult problems, \tsat{} still manages to solve most problems and maintains a comparably high success probability ($\ge 0.6\%$).
    For \walksat{}, we find two transitions to an extended intractable region which contains but is not limited to the critical line.
    Note that accidental solutions have $(r_1,r_2)\approx (0.46,0.41)$, and thus belong to a region that is comparably easy.
    For the point $A=(0.7,0.1)$ on and the point $B=(0.65,0.1)$ below the critical line, we compare the dependence on $N$ in \autoref{fig:psuccess_demo}.
    }
    \label{fig:fullspace_n1000_wsat_tsat}
\end{figure}
The aforementioned critical line of hardest problems is predicted to be located at $r_2=\frac{3}{2} - 2r_1$ for $r_3 \gtrsim 0.08$ \cite{sat.hard_random.generator.statmech.Weigt_2002}.
The line coincides with the simple condition that half the literals in the hidden solution should be true, and in the language of statistical physics this is equivalent to the statement that the local field (i.e. the terms linear in $\sigma_n$ in the Hamiltonian) should be zero for all variables.
The criterion about the starting point of the line at finite $r_3$ translates to the sudden onset of an extensive number of frozen variables in the solution (i.e. variables that take the same value in every solution) \cite{sat.hard_random.generator.statmech.Weigt_2002}.
We will later see in \autoref{sec:timescales} that this point corresponds to the location of a barrier in \gls{ctds} that takes exponential time to traverse.
\autoref{fig:fullspace_n1000_wsat_tsat} shows the average success probability $\langle p\rangle$ of individual trials for \walksat{} and our \tsat{} on problems generated using \algoref{alg:cnf} throughout the \gls{ctds} at $N=1000$ variables.
We find that for \walksat{}, $\langle p\rangle$ strongly depends on the position in \gls{ctds}: There exist extensive regions in which none of the $10^2$ problems could be solved within the $10^2$ trials per instance, and the transitions between easy and hard regions are quite sharp.
This indicates that the \gls{ctd} of the hidden solution indeed significantly impacts computational hardness for established \gls{sls} algorithms, and that problems even relatively far away from the critical line can be intractable for them.
\tsat{} on the other hand retains a comparably high success probability $\ge 0.6\%$ throughout \gls{ctds}, and the critical line is clearly distinguished as the location of the hardest problems.
In \autoref{sec:performance}, we discuss the scaling of $\langle p\rangle$ as a function of $N$ for the two highlighted points $A=(0.7,0.1)$ on the critical line and $B=(0.65,0.1)$ slightly below it.

\section{Targeting clause-type distributions}\label{sec:tsat}
In this section, we develop a \gls{sls} heuristic coined \textsc{Target-SAT} (\tsat{}) that can guide the search toward a given target point in \gls{ctds}.
Conceptually, specifying a target distribution amounts to selecting a sufficiently suitable picklock for the given problem instance out of (polynomially many) possible options.
On a more technical level, aside from the unique scoring function we also introduce the option to select already satisfied clauses as a valuable resource for overcoming barriers.

\subsection{Scoring via clause-type transitions}
In order to navigate the \gls{ctds}, our \tsat{} heuristic keeps track of all \emph{transitions} between type-$k$ clauses caused by a potential variable flip.
For 3-SAT, there are six possible transitions: $t_{k\rightarrow k+1}$ for $k=0,1,2$ and $t_{k\rightarrow k-1}$ for $k=1,2,3$ true literals.
Note that the breakcount relevant for \walksat{} corresponds to $b_n=t_{1\rightarrow 0}(x_n)$.
The change in the number of type-$k$ clauses upon flipping variable $x_n$, which we formally define as
\begin{align}
    \Delta_k(x_n) &= m_k(\bm{x} \text{ after flipping } x_n) - m_k(\bm{x}),\label{eqn:delta_k_of_x_n}
\end{align}
can be calculated from the transition counts via
\begin{align}
    \Delta_k &= t_{k+1\rightarrow k} - t_{k\rightarrow k+1} + t_{k-1\rightarrow k} - t_{k\rightarrow k-1}, \label{eqn:delta_k_from_tk}
\end{align}
where terms with invalid indices are neglected (affects $k=0,3$).
Now suppose that the current \gls{ctd} is $\bm{m}(\bm{x})$, and that the targeted \gls{ctd} is $\bm{t}$.
Then we define the distance vector $\bm{d}=\bm{t}-\bm{m}$, such that $d_0=-m_0=-E$, and introduce the scalar \emph{distance-to-target} as
\begin{align}
    \Delta_\mathrm{CTDS} &= \|\bm{d}\|_1 = \sum_{k=0}^3 | t_k - m_k |.\label{eqn:ctds_distance_to_target}
\end{align}
We seek to minimize this distance, and thus define the \emph{score} of variable $x_n$ as
\begin{align}
    s_n &= b_n + \sum_{k=1}^3 g_k |d_k - \Delta_k|. \label{eqn:tsat_score}
\end{align}
Here, $\bm{g}=(g_1,g_2,g_3)$ is a vector of couplings constants (i.e. algorithm parameters).
While one can certainly imagine different and also more complex scoring heuristics, \autoref{eqn:tsat_score} already incorporates two ideas: First, the score is proportional to the magnitude of the changes $\Delta_k$.
Second, by measuring the distance to $d_k$ we ensure that we prefer a move that actually brings us closer to the target (i.e. this deals with cases of overshooting).
One could consider introducing a dependence on $d_k$ in the couplings $g_k$, but we will see in \autoref{sec:timescales} that the bottleneck in the hardest problems lies in a highly localized barrier for which $g_k$ would remain approximately constant anyway.

To pick one of the three variables in a clause, we compute each score and then use a standard Boltzmann distribution with inverse temperature $\beta$ to select one, i.e. $P(x_n) \sim \text{e}^{-\beta s_n}$.
Because $s_n \sim d_n \sim M$, it is helpful to subtract the average score $\langle s\rangle$ from $s_n$ in the exponent to avoid numerical problems.
This is also why the score is effectively proportional to $\Delta_k$, as the $d_k$ is the same for every $s_n$ and thus cancels unless one is already very close to the target.

\subsection{Selecting satisfied clauses}
The scoring heuristic in \autoref{eqn:tsat_score} on its own can already significantly improve performance.
However, because of substantial energy barriers en-route to the target \gls{ctd} (cf.~\autoref{fig:caricature_3d}), we found that it can be very helpful to also assign a probability to select clauses that are already satisfied, but are of a type that we have too many of.
As an illustrative example, consider that we are at $\bm{m}=(5,25,15,5)$ but we want to reach $\bm{t}=(0,35,5,10)$.
Then the main obstruction is the abundance of type-2 clauses, which we can reduce by selecting one of them with a certain probability.
To achieve this, we define the (unnormalized) probability to select a type-$k$ clause as $p_k = -d_k$, truncating negative values to zero.
For the above example this would give $p_0=5$ and $p_2=10$, but it turns out that this is usually too skewed toward $p_2$.
To allow for some more control we scale the $p_k$ by a sigmoid,
\begin{align}
    p_k &\rightarrow p_k \cdot \frac{a_\mathrm{ampl}}{1 + \exp\left[ -a_\mathrm{sharpness}\left( \frac{p_k}{p_0}-1\right) \right]} \label{eqn:clause_pick_amplification}
\end{align}
for $k=1,2,3$.
The precise shape of this transition is found to be of minor quantitative importance and we always set $a_\mathrm{sharpness}=4$ in the following, while the amplification $a_\mathrm{ampl} \approx 0\dots 2$ is found to have a much bigger impact, and is thus adjusted as a model parameter.
For details on model parameters used in all benchmarked algorithms, see \appref{app:parameters}.
Finally, the probability to pick a type-$k$ clause is given by $P_k=p_k / \sum_{k=0}^3 p_k$.
The full variable selection process in \tsat{} is summarized in \algoref{alg:tsat}.

\begin{algorithm}[H]
\caption{Variable selection in \tsat{}}\label{alg:tsat}
\begin{algorithmic}[1]
\Require target $t_k$, and tables for $\Delta_k$ and $m_k$ (\autoref{eqn:delta_k_of_x_n} and \ref{eqn:delta_k_from_tk})
\State compute $p_k$ using \autoref{eqn:clause_pick_amplification} and normalize to $P_k$
\State $k\gets \text{random value based on probabilities } (P_0,\dots,P_3)$
\State clause $\gets$ random type-$k$ clause
\State compute the scores for variables in clause using \autoref{eqn:tsat_score}
\State $x_n \gets \text{random variable with probability} \sim\mathrm{e}^{-\beta s_n}$
\State \Return $x_n$
\end{algorithmic}
\end{algorithm}
Note that giving away the \gls{ctds} target information merely reduces the exponentially large search space by a polynomial amount since one could simply try out all the $\sim N^2$ possible target options (although more sophisticated search strategies are of course also conceivable).

\section{Performance Benchmarks}\label{sec:performance}
In this section, we zoom in on the system size scaling for the two points in \gls{ctds} highlighted in \autoref{fig:fullspace_n1000_wsat_tsat} to illustrate how the problem difficulty changes depending on the underlying target distribution.
We also compare the performance to \walksat{} and our recent \docsat{} \cite{sls.docsat.Schwardt_2025}.

\begin{figure}[!htp]
    \centering
    \includegraphics{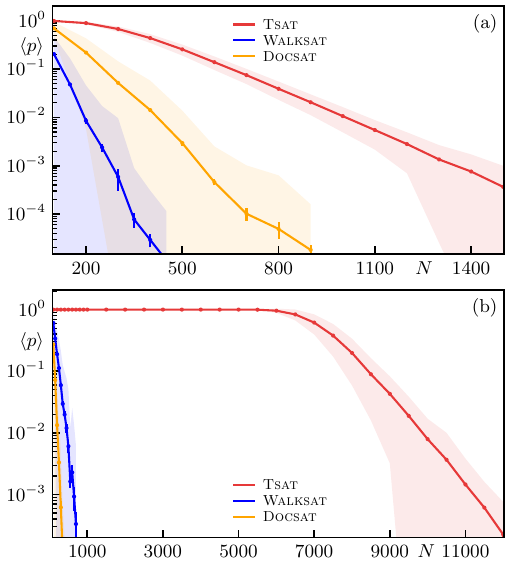}
    \caption{(a) Average success probability $\langle p \rangle$ of an individual trial for \walksat{}, \docsat{} and \tsat{} for $A=(0.7,0.1)$ ($10^3$ instances at each $N$ and $10^3$ trials).
    Consistent with the ratio of solved instances, \tsat{} shows the softest exponential decay and almost triples the accessible system size (runtime is $\sim 1/\langle p\rangle$).
    Also note that the variance is much smaller for \tsat{}, i.e. its performance is less instance dependent (shaded region marks the standard deviation).
    (b) At $B=(0.65,0.1)$, i.e. slightly below the critical line, the performance advantage of \tsat{} is even more striking.
    Interestingly, \tsat{} shows a rather sudden transition toward exponential scaling around $N\approx 6000$, which we investigate in \autoref{sec:timescales}.
    }
    \label{fig:psuccess_demo}
\end{figure}
For the point $A=(0.7,0.1)$ right on the critical line, which includes the most difficult known solvable problems, \autoref{fig:psuccess_demo}(a) shows that \tsat{} roughly triples the average accessible system size, which is remarkable given the exponential hardness of the problem.
The initially sub-exponential decay in the success probability $\langle p\rangle$ (implying sub-exponential runtime $\sim 1/\langle p\rangle$) eventually transitions to the expected exponential decay around $N\approx 500$.
At this system size, many problem instances are already intractable for other algorithms (including powerful complete solvers, cf.~\appref{app:comparison_others}).
While these problems are therefore evidently hard, they now look qualitatively easier thanks to the softer scaling of \tsat{} at these values of $N$.
Also note that the performance of e.g. \walksat{} is much more instance dependent, leading to a substantially larger variance:
In particular, the first unsolved instances in this dataset are already at a mere $N=100$ ($N=200$) variables for \walksat{} (\docsat{}), while only at $N=1000$ for \tsat{}.

For the point $B=(0.65,0.1)$ slightly below the critical line the success probability shown in \autoref{fig:psuccess_demo}(b) is again exponential for \walksat{} and \docsat{}, but here the latter performs significantly worse (cf.~\appref{app:comparison_others}).
\tsat{} meanwhile shows a very interesting transition from an apparently trivial to an exponentially hard regime around $N\approx 6000$, which is an order of magnitude beyond problem sizes accessible via other methods.
To understand this rather sudden change in scaling we have to take a closer look at the typical \tsat{} search trajectories through \gls{ctds}, which we analyze next.

\section{Distinguishing two types of complexity barriers}\label{sec:timescales}
In this section, we discuss in more detail the \tsat{} trajectories through \gls{ctds} as illustrated in \autoref{fig:caricature_3d}.
In summary, we find that the difficulty of the full problem is captured by two different types of complexity barriers, both of which are prominently featured in \autoref{fig:caricature_3d}.
One of these encompasses the \emph{final search} phase at very low energy in the vicinity of the targeted \gls{ctd}, and its size only shows a moderate dependence on the location of the hidden solution.
Since the neighborhood of this \gls{ctd} is an extensive subset of the full search space, it still contains exponentially many states, and therefore leads to a relatively soft but quite generic exponential wall, reflecting the NP-complete nature of the problem.
Interestingly, we also find that there exists a second barrier that strongly depends on the position in \gls{ctds} and represents a different kind of complexity.
It can dwarf the final search barrier by orders of magnitude in certain regions including the critical line, which allows us to elucidate why \walksat{} and other \gls{sls} algorithms have such a difficulty solving those problems.
To elaborate on the behavior of this additional barrier, in the following we analyze two representative points in \gls{ctds}, marked as $A$ and $B$ in \autoref{fig:fullspace_n1000_wsat_tsat}.

\begin{figure}[!htp]
    \centering
    \includegraphics{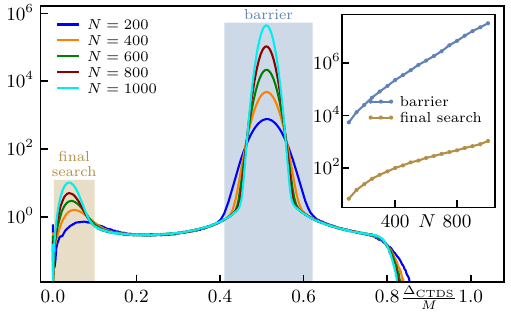}
    \caption{For varying size $N$, we run \tsat{} on $10^2$ instances at $A=(0.7,0.1)$ for $10^2$ trials each, with unlimited flips per trial (until a solution is found).
    The plot shows a histogram of the number of states $\bm{x}$ found at a certain distance-to-target (cf.~\autoref{eqn:ctds_distance_to_target}), with random initial states situated on the right and solutions near $\Delta_\mathrm{CTDS} \approx 0$ on the left.
    With increasing $N$, two peaks emerge and the search is dominated by the time it takes to traverse the barrier at $\Delta_\mathrm{CTDS}/M\approx 0.5$.
    Outside the shaded regions marking the peaks all lines coalesce, indicating that navigating the \gls{ctds} takes $\mathcal{O}(N)$ time in those regions.
    The inset shows the total number of states in the two peaks as a function of $N$.
    Both the barrier and final search time remain sub-exponential up to surprisingly large $N$ before transitioning to exponential scaling, which mirrors the success probability of \tsat{} in \autoref{fig:psuccess_demo}(a).
    }
    \label{fig:timescales_ctds}
\end{figure}

\subsection{Complexity barriers on the critical line}

First, consider the point $A=(0.7,0.1)$ on the critical line, in the vicinity of which problems are particularly difficult.
There, as shown in \autoref{fig:timescales_ctds}, the solution time is dominated by two diverging timescales:
As already discussed, the final search phase is expected to take exponentially long.
However, on the critical line a second timescale emerges due to a narrow but extensive barrier in \gls{ctds} around $\Delta_\mathrm{CTDS}/M \approx 0.5$.
This additional \emph{barrier time} is not only exponential, but also increases much faster than the timescale of the final search (cf.~\autoref{fig:timescales_ctds} inset), such that the runtime is dominated by the time it takes \tsat{} to reach the target region in \gls{ctds}.
This insight also explains more deeply why \walksat{} and other algorithms struggle so much with these problems, as they not only search in a part of the \gls{ctds} that does not contain the solution, but that is even separated from it by a large barrier rendering accidental success exceedingly improbable.

\begin{figure}[!htp]
    \centering
    \includegraphics{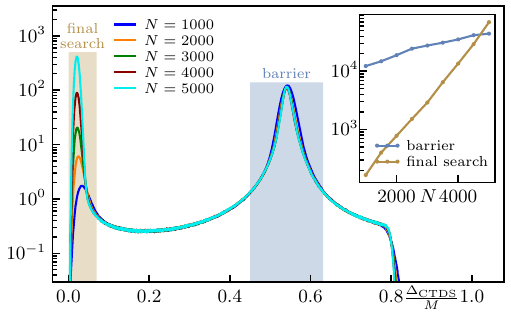}
    \caption{Setup as in \autoref{fig:timescales_ctds}, but at $B=(0.65,0.1)$.
    Away from the critical line the initial barrier appears to increase only linearly (note that the number of bins along the $x$-axis is $\propto N$, hence the coalescence of all lines).
    This explains why \tsat{} retains an average $\langle p \rangle \approx 1$ in \autoref{fig:psuccess_demo}(b) up to very large $N$; beyond $N\approx 5000$ the final search phase becomes the bottleneck (see crossing in the inset), and thus eventually inhibits solubility.
    }
    \label{fig:timescales_ctds_065}
\end{figure}
Another interesting observation in \autoref{fig:timescales_ctds} is the coalescence of all system sizes in the regions away from the two barrier peaks.
Because $\Delta_\mathrm{CTDS}$ is discrete and $\sim M$, the number of bins in the histogram is linear in $N$ (thus the integrated number of states is roughly $ \propto \text{peak  height}\times N$).
That all curves coalesce between distances of about $0.1$ to $0.4$ (similarly $0.65$ to $0.8$) implies that it takes \tsat{} a linear number of steps to reach the final search region before and after traversing the main barrier.
Thus, each individual step in \gls{ctds} toward the target in these regions actually takes only a constant time $\mathcal{O}(1)$ independent of system size.
Put differently, navigating the \gls{ctds} appears to rapidly transition from exponentially hard to negligible.

\subsection{Complexity off the critical line}

Moving slightly away from the critical line to the point $B=(0.65, 0.1)$, the runtime is now asymptotically dominated by the final search, because the barrier scales linearly (at least for the accessible system sizes).
However, although the final search eventually becomes the bottleneck, the break-even size is only around $N\approx 5000$, and thus completely out of reach of any other algorithm (cf.~\autoref{fig:timescales_ctds_065}).
We again stress that the $x$-axis has $\propto N$ many steps, which is why the barrier time in the inset still increases linearly despite the coalescing lines.

\begin{figure}[!htp]
    \centering
    \includegraphics{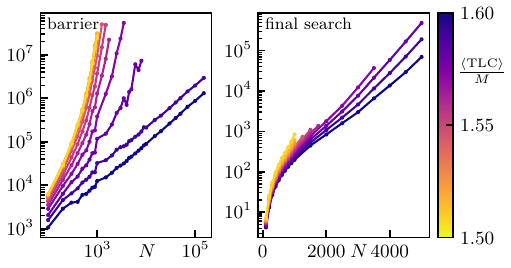}
    \caption{Total number of states in the barrier (left panel) and final search phase (right panel) as a function of system size $N$ for various equidistant average \gls{tlc} per clause of the hidden solution.
    The starting point is $B=(0.65,0.1)$ at $\langle \mathrm{TLC}\rangle / M = 1.6$ true literals per clause (cf.~\autoref{fig:timescales_ctds_065}) and decreasing the \gls{tlc} moves perpendicularly toward the critical line situated at $\langle \mathrm{TLC}\rangle / M = 1.5$.
    The final search always becomes exponentially expensive and in particular shows only a mild dependence on the position in \gls{ctds}.
    Sufficiently far away from the critical line the barrier appears to remain linear for all explored system sizes.
    Close to the line, the barrier dwarfs the final search by orders of magnitude for a given $N$ (see also~\autoref{fig:timescales_ctds}).
    }
    \label{fig:timescales_ctds_avg_tlc_line}
\end{figure}
Finally, in \autoref{fig:timescales_ctds_avg_tlc_line} we study how the scaling of the barrier and final search phase change as a function of distance from the critical line.
The deviation perpendicular to the critical line is encoded in the average number of true literals per clause,
\begin{align}
    \frac{\langle \mathrm{TLC}\rangle}{M} &= 0\cdot r_0 + 1\cdot r_1+2\cdot r_2+3\cdot r_3, \label{eqn:avg_tlc}
\end{align}
where the \acrfull{tlc} is the total number of true literals in a given state \cite{sls.docsat.Schwardt_2025}.
At the point $B=(0.65,0.1)$ we have $\langle \mathrm{TLC}\rangle / M=1.6$, where, as in \autoref{fig:timescales_ctds_065}, the final search asymptotically dominates the sub-exponential barrier.
As one approaches the point $(0.69,0.12)$ on the critical line with $\langle \mathrm{TLC}\rangle / M=1.5$, the barrier regains its sharp exponential scaling while the final search only shows a mild increase.
Note that the instance-to-instance variance of the barrier time appears to significantly increase around $\langle \mathrm{TLC}\rangle / M\approx 1.58$ (cf.~third line from the bottom in \autoref{fig:timescales_ctds_avg_tlc_line}).
This behavior may hallmark a criticality separating the exponential barrier from linear scaling.
As to what extent the parameter regime exhibiting exponential scaling may be further narrowed toward the critical line (or even overcome) by further algorithmic improvements remains an intriguing question for future work.

In the linear barrier regime, we note that we can abort the \tsat{} trials once the barrier is passed, making it possible to check the linear scaling well beyond $N=10^5$ without having to actually solve the problems (which would be prohibitively difficult due to the exponential final search).
Finally, we note that on the other side of the critical line, toward $\langle \mathrm{TLC}\rangle / M=1.4$, the overall situation is qualitatively similar as discussed above.

\section{Concluding discussion}\label{sec:outro}
We have shown how targeting a specific clause type distribution can efficiently guide stochastic local search toward a hidden solution, thus unlocking a vast range of random satisfiability problems, including a significant advance of the tractable systems sizes in the hardest known parameter regime.
This approach reveals and establishes \gls{ctds} as a natural phase space for solvable random 3-SAT instances beyond the comparably manageable random uniform problems.
The resulting \tsat{} algorithm largely improves on and generalizes our previous heuristic \docsat{} \cite{sls.docsat.Schwardt_2025} which in hindsight may be seen as targeting a lower dimensional cut through the full \gls{ctds} by simply reducing oversatisfied constraints (clauses with more than one true literal).

Besides the mere improvement in performance, our analysis also provides deep insights into the specific complexity of random 3-SAT instances in stochastic local search.
In particular, apart from isolated points in the explored \gls{ctds}, all instances, whether on the critical parameter line or not, exhibit a final search barrier in the vicinity to the solution that scales exponentially in problem size.
This highlights the generic exponential complexity of solving random 3-SAT problems in agreement with the exponential time hypothesis \cite{sat.np.complexity.eth.Impagliazzo_2001}.
Interestingly, the critical parameter line in \gls{ctds} is distinguished by an additional barrier far away from the solution that dominates the local search and represents the clear bottleneck for the tractable system sizes.
By contrast, away from the critical parameter line \tsat{} quickly overcomes this additional barrier in sub-exponential time, hence rendering the aforementioned final search the limiting structure, and explaining the rapid increase in solvable system sizes.

On a broader note, in this work we have made significant progress on solving hard random 3-SAT instances, and the principle of \tsat{} may readily be generalized to higher-dimensional \gls{ctds} so as to tackle $k$-SAT with $k>3$ as well as more generalized satisfiability problems with mixed clause lengths.
This being said, we would like to emphasize some key limitations and remaining open problems.
First, for instances that are not guaranteed to have a solution, proving their unsatisfiability remains a hard task for which complete solvers such as \cadical{} are leading.
Second, while the considered random problems are certainly most interesting in statistical physics and computational complexity theory, it remains to be seen as to what extent the principle behind \tsat{} may also lead to synergies in solving highly structured (or practical) problems.
There, the main challenge is to efficiently exploit a given non-generic structure to solve much larger system sizes (often $N \sim 10^6$), where the lead has so far also remained firmly in the realm of complete solvers.

\begin{acknowledgments}
    We acknowledge discussions with Tim Pokart and Yumin Hu as well as financial support from the German Research Foundation (DFG) through the Collaborative Research Centre (SFB 1143, project ID 247310070) and the Cluster of Excellence ctd.qmat (EXC 2147, project ID 390858490).
\end{acknowledgments}

\appendix

\section{Generating problems at fixed CTD}\label{app:cnf}
In order to generate random but satisfiable 3-SAT problem instances at a fixed position in \gls{ctds} we use the ideas from \cite{sat.hard_random.generator.statmech.Weigt_2002}.
Assuming that $\bm{x}=(1,\dots,1)$ is our hidden solution, we generate the requested number of $m_k=Mr_k$ type-$k$ clauses by choosing $k$ variables and $3-k$ negated variables as literals ($k=1,2,3$).
Finally, the solution is scrambled by negating each literal associated with a variable $x_n$ with probability $\frac{1}{2}$ for each $n=1,\dots,N$ (cf.~\algoref{alg:cnf}).
\begin{algorithm}[H]
\caption{Satisfiable random 3-SAT at fixed \gls{ctd}}\label{alg:cnf}
\begin{algorithmic}[1]
\Require $N$ variables, $M$ clauses and $r_1,r_2$
\State $\text{CNF} \gets \text{empty list of clauses}$
\For{$k=1,2,3$} \Comment{type-$k$ clauses}
\State $m_k \gets Mr_k$ \Comment{$m_3=M-m_1-m_2$}
\For{$m=1,\dots,m_k$} 
\State $\text{clause}\gets \text{3 random variables from }\{x_1,\dots,x_N\}$
\State randomly negate $(3-k)$ of the literals in clause
\State add clause to CNF
\EndFor
\EndFor
\For{$n=1,\dots,N$} \Comment{scramble the hidden solution}
\If{$\Call{RandomUniform}{0,1} < \frac{1}{2}$}
\State swap every occurrence of $x_n$ and $\bar{x}_n$ in $\text{CNF}$
\EndIf
\EndFor
\State \Return $\text{CNF}$
\end{algorithmic}
\end{algorithm}

\section{Comparison to other solvers}\label{app:comparison_others}
\begin{figure}[!htp]
    \centering
    \includegraphics{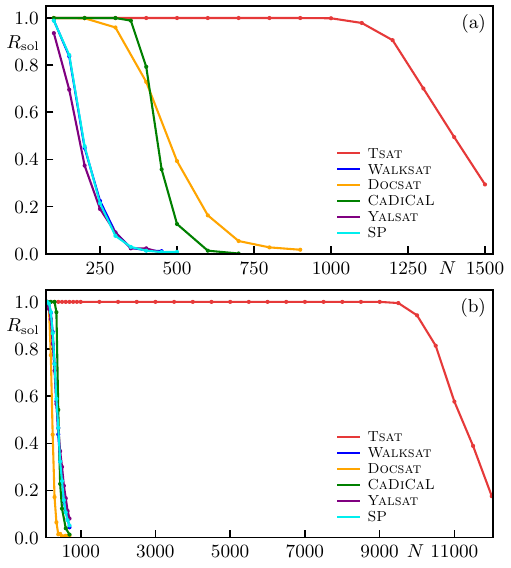}
    \caption{Ratio $R_\text{sol}$ of solved instances for $10^3$ problems and various algorithms (cf.~\appref{app:parameters} for parameters).
    (a) At the point $A=(0.7,0.1)$ on the critical line in \gls{ctds}, \walksat{}, \yalsat{} and \surveyprop{} do not perform well, but the complete solver \cadical{} benefits from the increased conflict-rate due to the larger than critical $\alpha=5$.
    \docsat{} still scales slightly better, while \tsat{} triples the accessible system size.
    (b) At the point $B=(0.65,0.1)$, \docsat{} performs poorly below the critical line (cf.~\autoref{fig:fullspace_dsat}).
    The other four algorithms are more similar, while \tsat{} stands out by solving all problems up to $N=10^4$.
    }
    \label{fig:singlespace_demo}
\end{figure}

Here, we compare \tsat{}'s performance to established algorithms, namely the very powerful complete solver \cadical{} \cite{solvers.cadical2.Biere_2024,solvers.kissat_et_al.SAT_comp2024.Biere_2024} based on conflict-driven clause learning, the \gls{sls} algorithm \yalsat{} \cite{sat.competition_2017.yalsat.Biere_2017}, and the message passing algorithm \surveyprop{} \cite{sls.sp.Braunstein_2005,sls.sp.bsp.ksat.Marino_2016}.
In \autoref{fig:singlespace_demo}, we show the ratio of solved problems $R_\text{sol}$ as a function of $N$ for (a) the point $A=(0.7,0.1)$ on the critical line and (b) the point $B=(0.65, 0.1)$ slightly below it (cf.~\autoref{fig:psuccess_demo}).
For both points, \surveyprop{} only manages to fix at most a handful of spins before reporting a paramagnetic state, at which stage the reduced problem is passed to \walksat{}.
Hence, the performance is almost identical to \walksat{} due to using the same number of total flips for both methods.
The performance of \cadical{} is mostly unaffected by the \gls{ctd} of the hidden solution, and the algorithm overall benefits slightly from the increased conflict-rate due to $\alpha>\alpha_\text{c}$.
However, due to the lack of structure in these random problems, it is still limited to $N\lesssim 500$.
Even for these hardest problems \tsat{} almost triples the accessible system size, while the advantage is well beyond an order of magnitude at the point $B$.
Our recent \docsat{} heuristic \cite{sls.docsat.Schwardt_2025} performs well at $A$, even solving some problems beyond \cadical{}, but works poorly at $B$, which we discuss next.

\begin{figure}[!htp]
    \centering
    \includegraphics{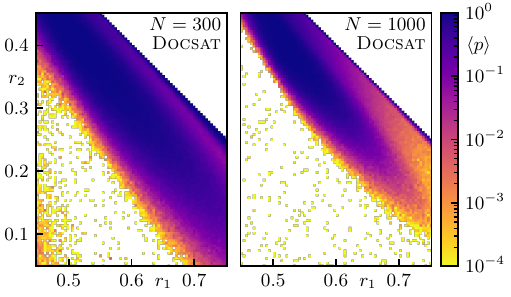}
    \caption{Average success probability $\langle p \rangle$ of \docsat{} for problems at various points in \gls{ctds} (cf.~\autoref{fig:fullspace_n1000_wsat_tsat}).
    The ratio of true literals in the hidden solution is $1 - \frac{2r_1+r_2}{3}$, while \docsat{} empirically operates near a ratio of $\approx 0.53$ (as opposed to $\approx 0.55$ for \walksat{}).
    This explains the poor performance below the critical line, but also why \docsat{} significantly outperforms \walksat{} in some regions.
    }
    \label{fig:fullspace_dsat}
\end{figure}

\docsat{} is based on a direct modification of \walksat{} with the purpose of guiding toward a desired global true literal count (\tsat{} draws inspiration from and generalizes this concept).
The concrete implementation of \docsat{} slightly favors variable flips that reduce the total number of true literals in the problem, essentially moving perpendicular to the critical line in \gls{ctds}.
On average, the ratio of true literals for \walksat{} is $\approx 0.55$ (which agrees well with the ratio for accidental solutions), while \docsat{} operates at a lower value of $\approx 0.53$.
Even though this may appear to be a small change, the effect on the performance throughout \gls{ctds} is dramatic as illustrated in \autoref{fig:fullspace_dsat}:
\docsat{} manages to improve drastically on \walksat{} in a part of the \gls{ctds}, in particular around the parameters studied in \cite{sls.docsat.Schwardt_2025}, while failing to solve most problems below the critical line (which includes the point $B$).
This is because the ratio of true literals in the hidden solution is $1 - \frac{2r_1+r_2}{3}$, which is $\frac{1}{2}$ on the critical line and larger than $\frac{1}{2}$ below it, such that the lower ratio targeted by \docsat{} is only superior to \walksat{} for problems situated close to or above the line.
When compared with \autoref{fig:fullspace_n1000_wsat_tsat}, one may view \docsat{} as quite complementary to \walksat{} in \gls{ctds}.
Importantly, the strong regime of \docsat{} contains the critical parameter line, even though the advantage to other solvers there decreases with increasing $r_1$ (decreasing $r_2$).

\section{Algorithm Parameters}\label{app:parameters}
\walksat{}: We set $p_\text{walk}=0.5$ and note that different values have a marginal impact on the results, as was also the case in our previous study \cite{sls.docsat.Schwardt_2025}.
Unless noted otherwise, we run $10^3$ trials on every problem instance with a cutoff of $300N$ flips per trial.

\docsat{}: As in \cite{sls.docsat.Schwardt_2025}, we set $p_\text{walk}=0.4$ and $r_\text{doc}=0.15$.
Flips and trials as for \walksat{}.

\tsat{}: We have optimized the parameters $\beta,a_\mathrm{ampl}$ and $\bm{g}$ for individual points from the \gls{ctds}, interpolating inbetween.
While these parameters yield substantially better results than a fixed global set, no fine-tuning is required in the sense that say $g_2=0.2$ is not much different from $g_2=0.1\dots 0.3$, while $g_2=0$ or $g_2=1$ may lead to significantly worse performance.
Flips and trials as for \walksat{}.

\surveyprop{}: We fix $1\,\%$ of variables per loop of the SP-iteration, with a limit of $10^3$ steps (that is rarely exhausted) and an error tolerance of $\epsilon=10^{-3}$ (same parameters as in \cite{sls.sp.Braunstein_2005}).
We use 25 full restarts of the \surveyprop{} elimination procedure with 40 \walksat{} trials after each (matching the total number of $10^3$ trials).
Because variables may be fixed to the wrong value (i.e. rendering an instance unsatisfiable), \surveyprop{} can perform worse than pure \walksat{} in some cases despite the increased runtime.
For problems near the critical line, performance is almost identical to \walksat{} because very few variables are fixed by \surveyprop{}.

\yalsat{}: Fewer full restarts appear to perform better for this algorithm, so we use $10^2$ trials with $3000N$ flips each.
Other parameters as in the default configuration.

\cadical{}: We have configured the solver to target satisfiable instances, which to our knowledge mainly impacts the restart interval.
The runtime scales slightly superlinear in the number of decisions, and a limit of $3000N$ yields runtime comparable to the other algorithms (but favoring \cadical{}).

\bibliography{bibliography}

@INPROCEEDINGS{sat.protein.Ollikainen_2009,
  author={Ollikainen, Noah and Sentovich, Ellen and Coelho, Carlos and Kuehlmann, Andreas and Kortemme, Tanja},
  booktitle={2009 IEEE/ACM International Conference on Computer-Aided Design - Digest of Technical Papers}, 
  title={SAT-based protein design}, 
  year={2009},
  volume={},
  number={},
  pages={128-135},
  doi={}
}

@article{sat.protein.optimization.Allouche_2014,
title = {Computational protein design as an optimization problem},
journal = {Artificial Intelligence},
volume = {212},
pages = {59-79},
year = {2014},
issn = {0004-3702},
doi = {https://doi.org/10.1016/j.artint.2014.03.005},
url = {https://www.sciencedirect.com/science/article/pii/S0004370214000332},
author = {David Allouche and Isabelle André and Sophie Barbe and Jessica Davies and Simon {de Givry} and George Katsirelos and Barry O'Sullivan and Steve Prestwich and Thomas Schiex and Seydou Traoré},
}

@article{sat.timetables.Matos_2021,
author={Matos, Gon{\c{c}}alo P. and Albino, Lu{\'i}s M. and Saldanha, Ricardo L. and Morgado, Ernesto M.},
title={Solving periodic timetabling problems with SAT and machine learning},
journal={Public Transport},
year={2021},
month={Oct},
day={01},
volume={13},
number={3},
pages={625-648},
issn={1613-7159},
doi={10.1007/s12469-020-00244-y},
url={https://doi.org/10.1007/s12469-020-00244-y}
}

@INPROCEEDINGS{sat.routing.system_on_a_chip.Zhukov_2020,
  author={Zhukov, Denis V. and Zheleznikov, Daniil A. and Zapletina, Mariya A.},
  booktitle={2020 IEEE Conference of Russian Young Researchers in Electrical and Electronic Engineering (EIConRus)},
  editor = {Shaposhnikov, S.},
  title={The Iterative SAT Based Approach to Detailed Routing for Reconfigurable System-on-a-Chip}, 
  year={2020},
  volume={},
  number={},
  pages={1905-1910},
  doi={10.1109/EIConRus49466.2020.9038929}
}

@INPROCEEDINGS{sat.routing.biochips.Yuh_2011,
  author={Yuh, Ping-Hung and Lin, Cliff Chiung-Yu and Huang, Tsung-Wei and Ho, Tsung-Yi and Yang, Chia-Lin and Chang, Yao-Wen},
  booktitle={International Workshop on System Level Interconnect Prediction}, 
  title={A SAT-based routing algorithm for cross-referencing biochips}, 
  year={2011},
  volume={},
  number={},
  pages={1-7},
  doi={10.1109/SLIP.2011.6135436}
}

@INPROCEEDINGS{sat.circuits.Nam_1999,
author = {Nam, Gi-Joon and Sakallah, Karem A. and Rutenbar, Rob A.},
booktitle = {International ACM Symposium on Field-Programmable Gate Arrays},
editor = {Kaptanoglu, Sinan and Trimberger, Steve},
title = {{Satisfiability-Based Layout Revisited: Detailed Routing of Complex FPGAs Via Search-Based Boolean SAT}},
year = {1999},
volume = {},
ISSN = {},
pages = {167-175},
doi = {10.1109/FPGA.1999.45},
url = {https://doi.ieeecomputersociety.org/10.1109/FPGA.1999.45},
publisher = {IEEE Computer Society},
address = {Los Alamitos, CA, USA},
month = {02}
}

@article{sat.circuits.FPGA_islands.Mukherjee_2015,
title = {SAT based solutions for detailed routing of island style FPGA architectures},
journal = {Microelectronics Journal},
volume = {46},
number = {8},
pages = {706-715},
year = {2015},
issn = {1879-2391},
doi = {https://doi.org/10.1016/j.mejo.2015.05.003},
url = {https://www.sciencedirect.com/science/article/pii/S0026269215001184},
author = {Shyamapada Mukherjee and Suchismita Roy},
}

@article{sat.pt.cavity.mc_data.alpha_crit.Lundow_2019,
  title = {Revisiting the cavity-method threshold for random 3-SAT},
  author = {Lundow, P. H. and Markstr\"om, K.},
  journal = {Phys. Rev. E},
  volume = {99},
  issue = {2},
  pages = {022106},
  numpages = {5},
  year = {2019},
  month = {02},
  publisher = {American Physical Society},
  doi = {10.1103/PhysRevE.99.022106},
  url = {https://link.aps.org/doi/10.1103/PhysRevE.99.022106}
}

@inproceedings{sat.pt.cavity.survey_prop.info_theo_threshold.Coja_2017,
author = {Coja-Oghlan, Amin and Krzakala, Florent and Perkins, Will and Zdeborova, Lenka},
title = {Information-theoretic thresholds from the cavity method},
year = {2017},
isbn = {9781450345286},
publisher = {Association for Computing Machinery},
address = {New York, NY, USA},
url = {https://doi.org/10.1145/3055399.3055420},
doi = {10.1145/3055399.3055420},
booktitle = {Proceedings of the 49th Annual ACM SIGACT Symposium on Theory of Computing},
editor = {Hatami, Hamed and McKenzie, Pierre and King, Valerie},
pages = {146–157},
numpages = {12},
location = {Montreal, Canada},
series = {STOC 2017}
}

@article{sat.pt.cavity.survey_prop.Coja_2017,
author = {Coja-Oghlan, Amin},
title = {Belief Propagation Guided Decimation Fails on Random Formulas},
year = {2017},
month = {02},
issue_date = {February 2017},
publisher = {Association for Computing Machinery},
address = {New York, NY, USA},
volume = {63},
number = {6},
issn = {0004-5411},
url = {https://doi.org/10.1145/3005398},
doi = {10.1145/3005398},
journal = {J. ACM},
articleno = {49},
numpages = {55}
}

@InProceedings{sat.survey_prop.random.analysis.Hetterich_2016,
  author =	{Hetterich, Samuel},
  title =	{Analysing Survey Propagation Guided Decimation on Random Formulas},
  booktitle =	{43rd International Colloquium on Automata, Languages, and Programming (ICALP 2016)},
  pages =	{65:1--65:12},
  IGNOREseries =	{Leibniz International Proceedings in Informatics (LIPIcs)},
  ISBN =	{978-3-95977-013-2},
  ISSN =	{1868-8969},
  year =	{2016},
  volume =	{55},
  editor =	{Chatzigiannakis, Ioannis and Mitzenmacher, Michael and Rabani, Yuval and Sangiorgi, Davide},
  publisher =	{Schloss Dagstuhl},
  IGNOREaddress =	{Dagstuhl, Germany},
  URL =		{https://drops.dagstuhl.de/entities/document/10.4230/LIPIcs.ICALP.2016.65},
  URN =		{urn:nbn:de:0030-drops-62197},
  doi =		{10.4230/LIPIcs.ICALP.2016.65}
}

@article{sat.pt.cavity.Mertens_2006,
author = {Mertens, Stephan and M\'{e}zard, Marc and Zecchina, Riccardo},
title = {Threshold values of random K-SAT from the cavity method},
year = {2006},
issue_date = {May 2006},
publisher = {John Wiley \& Sons, Inc.},
address = {USA},
volume = {28},
number = {3},
issn = {1042-9832},
journal = {Random Struct. Algorithms},
month = {05},
pages = {340–373},
numpages = {34}
}

@article{sat.pt.cavity.random.Mezard_2002,
author = {M. Mézard  and G. Parisi  and R. Zecchina },
title = {Analytic and Algorithmic Solution of Random Satisfiability Problems},
journal = {Science},
volume = {297},
number = {5582},
pages = {812-815},
year = {2002},
doi = {10.1126/science.1073287},
URL = {https://www.science.org/doi/abs/10.1126/science.1073287},
IGNOREeprint = {https://www.science.org/doi/pdf/10.1126/science.1073287}
}

@inproceedings{sat.NP_complete.Cook_1971,
author = {Cook, Stephen A.},
title = {The complexity of theorem-proving procedures},
year = {1971},
isbn = {9781450374644},
publisher = {Association for Computing Machinery},
address = {New York, USA},
url = {https://doi.org/10.1145/800157.805047},
doi = {10.1145/800157.805047},
booktitle = {Proceedings of the Third Annual ACM Symposium on Theory of Computing},
editor = {Harrison, Michael A. and Banerji, Ranan B. and Ullman, Jeffrey D.},
pages = {151–158},
numpages = {8},
location = {Shaker Heights, Ohio, USA},
series = {STOC '71}
}

@Article{sat.model_checking.survey.Prasad_2005,
author={Prasad, Mukul R. and Biere, Armin and Gupta, Aarti},
title={A survey of recent advances in SAT-based formal verification},
journal={International Journal on Software Tools for Technology Transfer},
year={2005},
month={04},
day={01},
volume={7},
number={2},
pages={156-173},
issn={1433-2787},
doi={10.1007/s10009-004-0183-4},
url={https://doi.org/10.1007/s10009-004-0183-4}
}

@InProceedings{sat.model_checking.verification.Gupta_2006,
author="Gupta, Aarti and Ganai, Malay K. and Wang, Chao",
editor="Bernardo, Marco and Cimatti, Alessandro",
title="SAT-Based Verification Methods and Applications in Hardware Verification",
booktitle="Formal Methods for Hardware Verification",
year="2006",
publisher="Springer Berlin Heidelberg",
address="Berlin, Heidelberg",
pages="108--143",
isbn="978-3-540-34305-9"
}

@article{sat.revolution.Fichte_2023,
author = {Fichte, Johannes K. and Berre, Daniel Le and Hecher, Markus and Szeider, Stefan},
title = {The Silent (R)evolution of SAT},
year = {2023},
issue_date = {June 2023},
publisher = {Association for Computing Machinery},
address = {New York, NY, USA},
volume = {66},
number = {6},
issn = {0001-0782},
url = {https://doi.org/10.1145/3560469},
doi = {10.1145/3560469},
journal = {Commun. ACM},
month = {05},
pages = {64–72},
numpages = {9}
}

@article{sat.review.advances.Alouneh_2019,
author = {Alouneh, Sahel and Abed, Sa’ed and Al Shayeji, Mohammad H. and Mesleh, Raed},
title = {A comprehensive study and analysis on SAT-solvers: advances, usages and achievements},
year = {2019},
issue_date = {Dec 2019},
publisher = {Kluwer Academic Publishers},
address = {USA},
volume = {52},
number = {4},
issn = {0269-2821},
url = {https://doi.org/10.1007/s10462-018-9628-0},
doi = {10.1007/s10462-018-9628-0},
journal = {Artif. Intell. Rev.},
month = {12},
pages = {2575–2601},
numpages = {27}
}

@Inbook{sat.practical.review.complete.Kullmann_2008,
author="Kullmann, Oliver",
editor="Creignou, Nadia and Kolaitis, Phokion G. and Vollmer, Heribert",
title="Present and Future of Practical SAT Solving",
bookTitle="Complexity of Constraints: An Overview of Current Research Themes",
year="2008",
publisher="Springer Berlin Heidelberg",
address="Berlin, Heidelberg",
pages="283--319",
isbn="978-3-540-92800-3",
doi="10.1007/978-3-540-92800-3_11",
url="https://doi.org/10.1007/978-3-540-92800-3_11"
}

@book{sat.handbook.Biere_2009,
  booktitle = {Handbook of Satisfiability},
  editor = {Biere, Armin and Heule, Marijn and van Maaren, Hans and Walsh, Toby},
  isbn = {978-1-58603-929-5},
  publisher = {IOS Press},
  series = {Frontiers in Artificial Intelligence and Applications},
  title = {Handbook of Satisfiability},
  url = {http://dblp.uni-trier.de/db/series/faia/faia185.html},
  volume = {185},
  year = {2009}
}

@InProceedings{sat.preprocessing.Een_2005,
author="E{\'e}n, Niklas and Biere, Armin",
editor="Bacchus, Fahiem and Walsh, Toby",
title="Effective Preprocessing in SAT Through Variable and Clause Elimination",
booktitle="Theory and Applications of Satisfiability Testing",
year="2005",
publisher="Springer Berlin Heidelberg",
address="Berlin, Heidelberg",
pages="61--75",
isbn="978-3-540-31679-4"
}

@inbook{sat.CDCL.solvers.Silva_2009,
title = "Conflict-driven clause learning SAT solvers",
author = "Joao Marques-Silva and Ines Lynce and Sharad Malik",
year = "2009",
doi = "10.3233/978-1-58603-929-5-131",
isbn = "9781586039295",
series = "Frontiers in Artificial Intelligence and Applications",
publisher = "IOS Press",
number = "1",
pages = "131--153",
booktitle = "Handbook of Satisfiability",
editor = "Biere, Armin and Heule, Marijn and van Maaren, Hans and Walsh, Toby",
address = "Netherlands",
edition = "1",
}

@article{sat.hard_random.generator.statmech.Weigt_2002,
  title = {Hiding Solutions in Random Satisfiability Problems: A Statistical Mechanics Approach},
  author = {Barthel, W. and Hartmann, A. K. and Leone, M. and Ricci-Tersenghi, F. and Weigt, M. and Zecchina, R.},
  journal = {Phys. Rev. Lett.},
  volume = {88},
  issue = {18},
  pages = {188701},
  numpages = {4},
  year = {2002},
  month = {04},
  publisher = {American Physical Society},
  doi = {10.1103/PhysRevLett.88.188701},
  url = {https://link.aps.org/doi/10.1103/PhysRevLett.88.188701}
}

@article{SLS.walksat.review.Hoos_2000,
author={Hoos, Holger H. and St{\"u}tzle, Thomas},
title={Local Search Algorithms for SAT: An Empirical Evaluation},
journal={Journal of Automated Reasoning},
year={2000},
month={05},
day={01},
volume={24},
number={4},
pages={421-481},
issn={1573-0670},
doi={10.1023/A:1006350622830},
url={https://doi.org/10.1023/A:1006350622830}
}

@inproceedings{sls.walksat.Selman_1994,
author = {Selman, Bart and Kautz, Henry A. and Cohen, Bram},
title = {Noise strategies for improving local search},
year = {1994},
isbn = {0262611023},
publisher = {American Association for Artificial Intelligence},
address = {USA},
booktitle = {Proceedings of the Twelfth National Conference on Artificial Intelligence (Vol. 1)},
editor = {Hayes-Roth, Barbara and Korf, Richard},
pages = {337–343},
numpages = {7},
location = {Seattle, Washington, USA},
series = {AAAI '94}
}

@Inbook{SLS.review.overview.Hoos_2015,
author="Hoos, Holger H. and St{\"u}tzle, Thomas",
editor="Kacprzyk, Janusz and Pedrycz, Witold",
title="Stochastic Local Search Algorithms: An Overview",
bookTitle="Springer Handbook of Computational Intelligence",
year="2015",
publisher="Springer Berlin Heidelberg",
address="Berlin, Heidelberg",
pages="1085--1105",
isbn="978-3-662-43505-2",
doi="10.1007/978-3-662-43505-2_54",
url="https://doi.org/10.1007/978-3-662-43505-2_54"
}

@article{SLS.WalkSAT.random_3sat.Fu_2020,
	author = {Huimin Fu and Yang Xu and Shuwei Chen and Jun Liu},
	title = {Improving WalkSAT for Random 3-SAT Problems},
	volume = {26},
	number = {2},
	year = {2020},
	doi = {10.3897/jucs.2020.013},
	publisher = {Journal of Universal Computer Science},
	issn = {0948-695X},
	pages = {220-243},
	URL = {https://doi.org/10.3897/jucs.2020.013},
	IGNOREeprint = {https://doi.org/10.3897/jucs.2020.013},
	journal = {JUCS - Journal of Universal Computer Science}
}

@software{walksat,
    author       = {Henry Kautz},
    title        = {WalkSAT Project -- version 57},
    year         = {2023},
    month        = {11},
    version      = {57},
    url          = {https://gitlab.com/HenryKautz/Walksat}
}

@inproceedings{sat.hard_regime.Cheeseman_1991,
author = {Cheeseman, Peter and Kanefsky, Bob and Taylor, William M.},
title = {Where the really hard problems are},
year = {1991},
isbn = {1558601600},
publisher = {Morgan Kaufmann Publishers Inc.},
address = {San Francisco, CA, USA},
booktitle = {Proceedings of the 12th International Joint Conference on Artificial Intelligence - Volume 1},
editor = {Mylopoulos, John and Reiter, Ray},
pages = {331–337},
numpages = {7},
location = {Sydney, New South Wales, Australia},
series = {IJCAI'91}
}

@inproceedings{sat.hard_regime.distributions.Mitchell_1992,
author = {Mitchell, David and Selman, Bart and Levesque, Hector},
title = {Hard and easy distributions of SAT problems},
year = {1992},
isbn = {0262510634},
publisher = {AAAI Press},
booktitle = {Proceedings of the Tenth National Conference on Artificial Intelligence},
editor = {Rosenbloom, Paul and Szolovits, Peter},
pages = {459–465},
numpages = {7},
location = {San Jose, California},
series = {AAAI'92}
}

@InProceedings{solvers.cadical2.Biere_2024,
author="Biere, Armin and Faller, Tobias and Fazekas, Katalin and Fleury, Mathias and Froleyks, Nils and Pollitt, Florian",
editor="Gurfinkel, Arie and Ganesh, Vijay",
title="CaDiCaL 2.0",
booktitle="Computer Aided Verification",
year="2024",
publisher="Springer Nature Switzerland",
address="Cham",
pages="133--152",
isbn="978-3-031-65627-9"
}

@inproceedings{solvers.kissat_et_al.SAT_comp2024.Biere_2024,
  author       = {Armin Biere and Tobias Faller and Katalin Fazekas and Mathias Fleury and Nils Froleyks and Florian Pollitt},
  title	       = {{CaDiCaL}, {Gimsatul}, {IsaSAT} and {Kissat} Entering the {SAT Competition 2024}},
  editor       = {Heule, Marijn and Iser, Markus and J{\"a}rvisalo, Matti and Suda, Martin},
  bookTitle    = {Proc.~of {SAT Competition} 2024 -- Solver, Benchmark and Proof Checker Descriptions},
  volume       = {B-2024-1},
  IGNOREseries       = {Department of Computer Science Report Series B},
  publisher    = {University of Helsinki},
  year	       = {2024},
  pages	       = {8-10},
}

@inproceedings{sat.competition_2017.yalsat.Biere_2017,
title = "CaDiCaL, Lingeling, Plingeling, Treengeling, YalSAT Entering the SAT Competition 2017",
author = "Armin Biere",
year = "2017",
month = "09",
volume = "B-1",
IGNOREseries = "Department of Computer Science Series of Publications B",
publisher = "University of Helsinki",
pages = "14--15",
bookTitle = "Proceedings of SAT Competition 2017 - Solver and Benchmark Descriptions",
editor = "Tom{\'a}{\v s}, Tomas and Heule, Marijn and J{\"a}rvisalo, Matti",
}

@article{np_complete.ising_formulations.lucas_2014, 
author = {Lucas, Andrew},
year = {2014},
month = {02},
pages = {5},
title = {Ising formulations of many NP problems},
volume = {2},
journal = {Frontiers in Physics},
doi = {10.3389/fphy.2014.00005},
issn={2296-424X}
}

@book{np_complete.guide.computers_intractibility.Garey_1990,
author = {Garey, Michael R. and Johnson, David S.},
title = {Computers and Intractability; A Guide to the Theory of NP-Completeness},
year = {1990},
isbn = {0716710455},
publisher = {W. H. Freeman \& Co.},
address = {USA}
}

@inbook{np_complete.reducibility.combinatoric_problems.Karp_1972,
author="Karp, Richard M.",
editor="Miller, Raymond E. and Thatcher, James W. and Bohlinger, Jean D.",
title="Reducibility among Combinatorial Problems",
IGNOREbookTitle="Complexity of Computer Computations: Proceedings of a symposium on the Complexity of Computer Computations, held March 20--22, 1972, at the IBM Thomas J. Watson Research Center, Yorktown Heights, New York, and sponsored by the Office of Naval Research, Mathematics Program, IBM World Trade Corporation, and the IBM Research Mathematical Sciences Department",
bookTitle="Proceedings of a symposium on the Complexity of Computer Computations",
year="1972",
publisher="Springer US",
address="Boston, MA",
pages="85--103",
isbn="978-1-4684-2001-2",
doi="10.1007/978-1-4684-2001-2_9",
url="https://doi.org/10.1007/978-1-4684-2001-2_9"
}

@article{np_complete.comp_complexity.ising_spinglass.Barahona_1982,
doi = {10.1088/0305-4470/15/10/028},
url = {https://dx.doi.org/10.1088/0305-4470/15/10/028},
year = {1982},
month = {10},
publisher = {IOP Publishing},
volume = {15},
number = {10},
pages = {3241},
author = {F Barahona},
title = {On the computational complexity of Ising spin glass models},
journal = {Journal of Physics A: Mathematical and General},
}

@inproceedings{ai.prob_reasoning.maxsat.Park_2002,
author = {Park, James D.},
title = {Using weighted MAX-SAT engines to solve MPE},
year = {2002},
isbn = {0262511290},
publisher = {American Association for Artificial Intelligence},
address = {USA},
booktitle = {Eighteenth National Conference on Artificial Intelligence},
pages = {682–687},
numpages = {6},
location = {Edmonton, Alberta, Canada}
}

@article{ai.maxsat.maxsolver.Xing_2005,
title = {MaxSolver: An efficient exact algorithm for (weighted) maximum satisfiability},
journal = {Artificial Intelligence},
volume = {164},
number = {1},
pages = {47-80},
year = {2005},
issn = {0004-3702},
doi = {https://doi.org/10.1016/j.artint.2005.01.004},
url = {https://www.sciencedirect.com/science/article/pii/S0004370205000160},
author = {Zhao Xing and Weixiong Zhang},
}

@book{ai.constraint_processing.Dechter_2003,
  author = {Dechter, Rina},
  publisher = {Morgan Kaufmann},
  title = {Constraint Processing},
  year = {2003}
}

@inproceedings{ai.MMP.MarquesSilva_2013,
author = {Marques-Silva, Joao and Janota, Mikol\'{a}\v{s} and Belov, Anton},
title = {Minimal Sets over Monotone Predicates in Boolean Formulae},
year = {2013},
isbn = {9783642397981},
publisher = {Springer-Verlag},
address = {Berlin, Heidelberg},
booktitle = {Proceedings of the 25th International Conference on Computer Aided Verification - Volume 8044},
editor = {Sharygina, Natasha and Veith, Helmut},
pages = {592–607},
numpages = {16},
location = {Saint Petersburg, Russia},
series = {CAV 2013}
}

@article{sls.sp.Braunstein_2005,
author = {Braunstein, A. and M\'{e}zard, M. and Zecchina, R.},
title = {Survey propagation: An algorithm for satisfiability},
year = {2005},
issue_date = {September 2005},
publisher = {John Wiley \& Sons, Inc.},
address = {USA},
volume = {27},
number = {2},
issn = {1042-9832},
journal = {Random Struct. Algorithms},
month = {09},
pages = {201–226},
numpages = {26}
}

@article{sls.sp.bsp.ksat.Marino_2016,
author={Marino, Raffaele and Parisi, Giorgio and Ricci-Tersenghi, Federico},
title={The backtracking survey propagation algorithm for solving random K-SAT problems},
journal={Nature Communications},
year={2016},
month={10},
day={03},
volume={7},
number={1},
pages={12996},
issn={2041-1723},
doi={10.1038/ncomms12996},
url={https://doi.org/10.1038/ncomms12996}
}

@article{sat.np.complexity.eth.Impagliazzo_2001,
title = {On the Complexity of k-SAT},
journal = {Journal of Computer and System Sciences},
volume = {62},
number = {2},
pages = {367-375},
year = {2001},
issn = {0022-0000},
doi = {https://doi.org/10.1006/jcss.2000.1727},
url = {https://www.sciencedirect.com/science/article/pii/S0022000000917276},
author = {Russell Impagliazzo and Ramamohan Paturi}
}

@article{sat.pt.random.two_moments.Achlioptas_2006,
author = {Achlioptas, Dimitris and Moore, Cristopher},
title = {Random k‐SAT: Two Moments Suffice to Cross a Sharp Threshold},
journal = {SIAM Journal on Computing},
volume = {36},
number = {3},
pages = {740-762},
year = {2006},
doi = {10.1137/S0097539703434231},
URL = {https://doi.org/10.1137/S0097539703434231},
IGNOREeprint = {https://doi.org/10.1137/S0097539703434231}
}

@article{sls.docsat.Schwardt_2025,
author = {Joachim Schwardt and Jan Carl Budich },
title = {Advancing stochastic 3-SAT solvers by dissipating oversatisfied constraints},
journal = {Proceedings of the National Academy of Sciences},
volume = {122},
number = {46},
year = {2025},
URL = {https://www.pnas.org/doi/abs/10.1073/pnas.2517297122},
doi = {10.1073/pnas.2517297122},
IGNOREeprint = {https://www.pnas.org/doi/pdf/10.1073/pnas.2517297122},
pages = {e2517297122},
}

@book{optimization.algorithms.physics.Hartmann_2006,
  title={New Optimization Algorithms in Physics},
  author={Hartmann, A.K. and Rieger, H.},
  isbn={9783527604579},
  url={https://books.google.de/books?id=zvn11pIJPsIC},
  year={2006},
  publisher={Wiley}
}

@article{spinglass.sk_model.infinite_order_pars.Parisi_1979,
  title = {Infinite Number of Order Parameters for Spin-Glasses},
  author = {Parisi, G.},
  journal = {Phys. Rev. Lett.},
  volume = {43},
  issue = {23},
  pages = {1754--1756},
  numpages = {0},
  year = {1979},
  month = {12},
  publisher = {American Physical Society},
  doi = {10.1103/PhysRevLett.43.1754},
  url = {https://link.aps.org/doi/10.1103/PhysRevLett.43.1754}
}

@article{spinglass.sk_model.solution_sequences.Parisi_1980,
doi = {10.1088/0305-4470/13/4/009},
url = {https://doi.org/10.1088/0305-4470/13/4/009},
year = {1980},
month = {apr},
publisher = {IOP},
volume = {13},
number = {4},
pages = {L115},
author = {G Parisi},
title = {A sequence of approximated solutions to the S-K model for spin glasses},
journal = {Journal of Physics A: Mathematical and General}
}

@article{eo.algorithm.Boettcher_2001,
  title = {Optimization with Extremal Dynamics},
  author = {Boettcher, Stefan and Percus, Allon G.},
  journal = {Phys. Rev. Lett.},
  volume = {86},
  issue = {23},
  pages = {5211--5214},
  numpages = {0},
  year = {2001},
  month = {06},
  publisher = {American Physical Society},
  doi = {10.1103/PhysRevLett.86.5211},
  url = {https://link.aps.org/doi/10.1103/PhysRevLett.86.5211}
}

@article{ea.spinglass.perf_match.Kasteleyn_cities.Thomas_2007,
  title = {Matching Kasteleyn cities for spin glass ground states},
  author = {Thomas, Creighton K. and Middleton, A. Alan},
  journal = {Phys. Rev. B},
  volume = {76},
  issue = {22},
  pages = {220406},
  numpages = {4},
  year = {2007},
  month = {12},
  publisher = {American Physical Society},
  doi = {10.1103/PhysRevB.76.220406},
  url = {https://link.aps.org/doi/10.1103/PhysRevB.76.220406}
}

@article{ea.spinglass.perf_match.Bieche_1980,
doi = {10.1088/0305-4470/13/8/005},
url = {https://doi.org/10.1088/0305-4470/13/8/005},
year = {1980},
month = {08},
publisher = {IOP},
volume = {13},
number = {8},
pages = {2553},
author = {L Bieche and J P Uhry and R Maynard and R Rammal},
title = {On the ground states of the frustration model of a spin glass by a matching method of graph theory},
journal = {Journal of Physics A: Mathematical and General}
}

@inbook{sat.complexity.around_alpha_c.Coarfa_2000,
author = {Coarfa, Cristian and Demopoulos, Demetrios and San Miguel Aguirre, Alfonso and Subramanian, Devika and Vardi, Moshe},
year = {2000},
publisher = {Springer Berlin},
bookTitle = {Principles and Practice of Constraint Programming},
editor = {Dechter, Rina},
month = {01},
pages = {143-159},
title = {Random 3-SAT: The Plot Thickens},
volume = {8},
isbn = {978-3-540-41053-9},
journal = {Constraints},
doi = {10.1007/3-540-45349-0_12}
}

@article{sat.qubo.comb.optimization.landscapes.deep_minima.Dobrynin_2024,
  title = {Energy landscapes of combinatorial optimization in Ising machines},
  author = {Dobrynin, Dmitrii and Renaudineau, Adrien and Hizzani, Mohammad and Strukov, Dmitri and Mohseni, Masoud and Strachan, John Paul},
  journal = {Phys. Rev. E},
  volume = {110},
  issue = {4},
  pages = {045308},
  numpages = {19},
  year = {2024},
  month = {10},
  publisher = {American Physical Society},
  doi = {10.1103/PhysRevE.110.045308},
  url = {https://link.aps.org/doi/10.1103/PhysRevE.110.045308}
}

\end{document}